\documentclass{svjour2}                    
\smartqed  
\usepackage{graphicx}
\usepackage{amsmath}
\renewcommand\Re{\operatorname{Re}}

\begin{document}

\title{A Signal Processing Model of Quantum Mechanics
}


\author{Chris Thron         \and
        Johnny Watts 
}


\institute{C. Thron \at
              Department of Mathematics, Texas A\&M University - Central Texas \\
              Tel.: +1-254-5195776\\
              \email{thron@ct.tamus.edu}           
           \and
           J. Watts \at
              Tel.: +1-254-9314456\\
}

\date{Received: date / Accepted: date}

\maketitle

\begin{abstract}
This paper develops a deterministic model of quantum mechanics as an accumulation-and-threshold process. The model arises from an analogy with signal processing in wireless communications. Complex wavefunctions are interpreted as expressing the amplitude and phase information of a modulated carrier wave. Particle transmission events are modeled as the outcome of a process of signal accumulation that occurs in an extra (non-spacetime) dimension.  

Besides giving a natural interpretation of the wavefunction and the Born rule,  the model accommodates the collapse of the wave packet and other quantum paradoxes such as EPR and the Ahanorov-Bohm effect. The model also gives a new perspective on the `relational' nature of quantum mechanics: that is, whether the wave function of a physical system is ``real" or simply reflects the observer's partial knowledge of the system.  We simulate the model for a 2-slit experiment, and indicate possible deviations of the model's predictions from conventional quantum  mechanics. We also indicate how the theory may be extended to a field theory. 

\end{abstract}

\section{Wavefunction analogy in wireless communications}
\label{sec:wireless}
Several physical systems are characterized by a process of accumulation (of energy, charge, etc.), which leads to an activation event once the accumulation attains a certain threshold. Examples of such systems include lightning and nerve impulse transmission. In many cases the accumulation process is described in terms of a continuous field, while attaining the threshold triggers a discrete event. This simultaneous presence of discrete and continuous aspects is reminiscent of quantum mechanics.  

Signal acquisition in wireless digital communications also follows this same general pattern. Consider a mobile receiver moving randomly within a region in which a modulated carrier wave is broadcast. The carrier wave is modulated both in amplitude and phase. In order to detect the broadcasted signal, the receiver accumulates its received signal until a detection threshold is reached. We shall construct a mathematical model of a system, in which that the location where detection occurs obeys a probability distribution reminiscent of the quantum wavefunction.
In our model, the wireless signal has the following characteristics: 
\begin{itemize}
\vspace{1mm}
\item The carrier frequency is  $\omega$, so that the signal has the general mathematical form $A(\mathbf{r},t)\sin(\omega t+ \phi(\mathbf{r},t))$. Such  a signal is commonly represented by its ``complex amplitude" $A(\mathbf{r},t) e^{i\phi(\mathbf{r},t)}$. 
\vspace{1mm}
\item The transmitted signal (at the transmitter) has constant complex amplitude over time intervals of length $\delta$, where $\delta >> 2\pi / \omega$ ($\delta$ is called the ``chip width'' in digital communications (\cite{Proakis})). The probability distribution of complex amplitudes is Gaussian, so that  real and imaginary parts are independent, identically distributed (i.i.d)  standard normal random variables with mean 0  and variance 1.
\vspace{1mm}
\item The ratio of field amplitude to transmitted signal amplitude (denoted by $\psi(\mathbf{r}))$ depends on the field location $\mathbf{r}$, but is independent of time. For mathematical simplicity, we assume that the ratio assumes one of a finite set of complex values $\lbrace \psi_1,\ldots \psi_K \rbrace$; and that within the (finite) region of interest,  the sets $\lbrace \mathbf{r} | \psi(\mathbf{r}) = \psi_k~ (k=1,\ldots K) \rbrace$  all have equal area (see Figure 1).
\end{itemize}
\begin{figure}
  \includegraphics{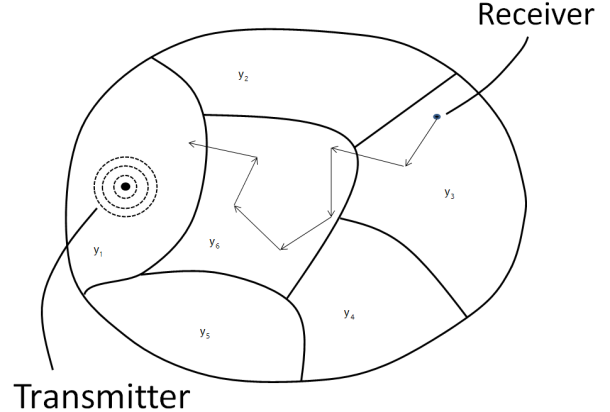}
\caption{Wireless communication model}
\label{fig:1} 
\end{figure}
The receiver has the following characteristics:
\begin{itemize}
\vspace{1mm}
\item  The receiver consists of an oscillating circuit with natural frequency $\omega$, which is driven by the signal field at the receiver's current location.
\vspace{1mm}
\item  The receiver moves slowly enough so that its field amplitude does not change significantly over time intervals of length $M\delta$, where $M$ is an integer $>> 1$.
\vspace{1mm}
\item  The receiver moves in such a way that its position uniformly samples the entire region of interest (for instance, by random walk). 
\vspace{1mm}
\item Our mathematical proof (see Appendix) requires that the receiver's fields over the time intervals $(m_1,m_1+1)M\delta$ and $(m_2,m_2+1)M\delta$ are statistically independent whenever $ m_1 \neq m_2$. Strictly speaking, a receiver moving under random walk will not satisfy this condition: instead, the receiver would have to make uniformly-distributed random jumps at times $M\delta, 2M\delta, \ldots $  A rigorous treatment with random-walk motion would require a more careful analysis. 
\vspace{1mm}
\item  The receiver detects the signal when the receiver's amplitude exceeds a fixed threshold $\Theta$.
\end{itemize}	
Note this simple model does not include any effects from polarization, propagation delay or Doppler phase shifting.
With the above assumptions, the field at receiver position r can be expressed (with the aid of complex amplitudes) as follows:
\begin{equation}
A(\mathbf{r},t) = \Re[ \psi(\mathbf{r})\cdot \nu_{\lceil t/ \delta \rceil} \cdot e^{i\omega t}]  \label{eq:1.1}
\end{equation}
where
\begin{description}
\item[]   $\psi(\mathbf{r})$  = (field amplitude at $\mathbf{r}$) / (signal amplitude at transmitter)
\vspace{1mm}
\item[]   $\nu_{\lceil t/\delta \rceil}$  may be written as $\nu_n=\alpha_n+i\beta_n$, where $\alpha_n,\beta_n$ are i.i.d. standard normal random variables. 
\vspace{1mm}
\item[]  $\lceil x \rceil$ denotes the ``ceiling" function, i.e. next largest integer greater than $x$.
\vspace{1mm}
\end{description}
We now suppose that the trajectory of the receiver is given by the function $\mathbf{r}(t)$.  It follows that the equation for the amplitude $x(t)$ of the driven oscillator is:
\begin{equation}
	x''+ \omega^2x = A(\mathbf{r}(t),t)    \label{eq:1.2}
\end{equation}
This equation  may be expressed as the real part of the complex equation
\begin{equation}
z'' + \omega^2z = \psi(\mathbf{r}(t))\cdot \nu_{\lceil t/ \delta \rceil} \cdot e^{i\omega t}   \label{eq:1.3}
\end{equation}
The solution of (\ref{eq:1.3}) which satisfies $z(0) = z'(0) = 0$ is
\begin{equation}
 z(t)=  \frac{-i}{2\omega} \int_0^t \psi(\mathbf{r}(u))\cdot \nu_{\lceil u/\delta \rceil} du \cdot e^{i\omega t} + \frac{i}{2 \omega} \int_0^t \psi(\mathbf{r}(u)) \cdot \nu_{\lceil u/ \delta \rceil}  \cdot e^{2i \omega u} du \cdot e^{-i\omega t}    \label{eq:1.4}
\end{equation}
According to our assumptions, the factor $e^{2i \omega u}$ in the second integrand oscillates rapidly compared to the rest of the integrand, which causes the second integral to be negligible compared to the first.
The model assumptions imply that $\psi(\mathbf{r}(u))$ can be treated as constant over time intervals of length $M\delta$.  Using the notation $\Psi_{\lceil u/(M\delta) \rceil} \equiv \psi(\mathbf{r}(u))$, we have: 
\begin{equation}
z(t) \approx  \frac{-i\delta}{2\omega} \sum_{n=1}^{\lceil t/\delta \rceil} \Psi_{\lceil n/M \rceil} \cdot \nu_n \cdot e^{i \omega t} \label{eq:1.5}
\end{equation}
The oscillation at time $N \delta$ has complex amplitude $(-i\delta/2\omega) \cdot S(N)$, where
\begin{equation}
S(N) \equiv \sum_{n=1}^N \Psi_{\lceil n/M \rceil} \cdot \nu_n   \label{eq:1.6}
\end{equation}
According to the model assumptions, each $\Psi_j$ is one of the values $\lbrace\psi_1, \ldots \psi_K \rbrace$.  Define the random variable $\kappa(m)$ to be the index $k$ corresponding to random variable $\Psi_m$: that is
\begin{equation}
	\kappa(m) \equiv \lbrace k | \Psi_m = \psi_k \rbrace   \label{eq:1.7}
\end{equation}
Define $N_\Theta$  as the time index at which $|S(N)|$ first passes a given threshold $\Theta$:
\begin{equation}
	N_\Theta \equiv \min_N⁡ \lbrace N| \sim |S(N)| \ge \Theta \rbrace   \label{eq:1.8}
\end{equation}	 
Our goal is to evaluate the probability distribution of $\kappa (\ldots )$ corresponding to the first passing of the threshold:
\begin{equation}
\Pr\left[\kappa \left( \left \lceil N_{\Theta} / M \right \rceil \right)=k\right] \qquad k =1,\ldots ,K,    \label{eq:1.9}
\end{equation}
This corresponds to the probability distribution of the value of field $\psi$ at the location of detection. 
It is a well-known fact in signal processing that the rate of accumulation of a random signal is proportional to the signal power, which is in turn proportional to the squared signal amplitude (\cite{Proakis}). It stands to reason that given a signal that assumes different power levels at different times but with equal probabilities, the chance of the accumulated signal passing a fixed threshold while at a certain power level should be proportional to that power level. This is in fact the case; in the Appendix we prove that 
\begin{equation}
\Pr\left[\kappa \left( \left \lceil N_{\Theta} /M \right \rceil \right)=k\right] \propto |\psi_k |^2,	\qquad k =1,\ldots ,K,      \label{eq:1.10}
\end{equation}
which are exactly the Born probabilities for the spatial wavefunction $\psi$.
\section{Single quantum detection event model}
\label{sec:single}
In this section, we present a model (based on the model in the previous section) that explains quantum detection probabilities. The model is discretized for conceptual clarity and computational tractability; it is fairly straightforward to see how the model could be taken to a continuous limit.

We emphasize that the probability distribution in the previous model arose from the outcome of a process. The process involved sampling the entire region of potential detection before the actual detection was made. This representative sampling was necessary in order for the field strengths to translate into relative probabilities. We want similar characteristics for the quantum process.
 
In our previous model the process variable is time; this was appropriate because we were only concerned about the spatial position of the receiver at the moment of detection. However, in quantum mechanics, we are concerned about the location of detection events within space-time. It is impossible to have a process that unfolds in time that at the same time samples all space-time locations before determining the detection location. For this reason, it is necessary to introduce a new process variable, so that the process of signal accumulation takes place in a non-observable dimension which we will call the $a$-dimension.

Our wireless communications model had a physical receiver which moves within the state space of possible detection locations. Quantum detection (say of a particle on a screen) does not appear to have any corresponding receiver. We therefore introduce the notion of a \emph{detectron}, which plays the same role as the receiver in our previous model. The accumulation takes place as the detectron jumps around and uniformly samples the set of all potential detection locations. This ``jumping around'' takes place in the $a$-dimension; for fixed $a$, the detectron's space-time location is fixed. We emphasize that the detectron is a mathematical construct, and should not be considered as a physical particle; we will say more about the physical nature of detectrons in Section~\ref{sec:extension}.

We also postulate a carrier wave that oscillates as a function of $a$ (\emph{not} as a function of time) having the mathematical form $\sin\omega a$.  The frequency $\omega$ is unknown, and does not correspond to any measurable quantity in space-time.  The signal has the following characteristics:
\begin{itemize}
\vspace{1mm}
\item  The signal has constant complex amplitude over $a$-intervals of length $\delta$, where $\delta >> 2\pi/\omega$. The distribution of complex amplitudes is mean-zero Gaussian, with i.i.d. standard normal real and imaginary parts;
\vspace{1mm}
\item  The signal is multiplied by a complex field amplitude $\psi(\mathbf{r},t)$ which is independent of $a$. For mathematical simplicity, we assume that the amplitude assumes one of a finite set of complex values $\lbrace \psi_1, \ldots \psi_K \rbrace$; and that within the space-time confines of the detector, the sets $\lbrace \mathbf{r},t | \psi(\mathbf{r},t) = \psi_k \rbrace~ (k=1,\ldots K)$  all have equal 4-volume.
\end{itemize}
The detectron has the following characteristics:
\begin{itemize}
\vspace{1mm}
\item	Associated with the detectron is an oscillator (which varies sinusoidally with $a$) with natural frequency $\omega$, which is driven by the signal field at the detectron's current space-time location;
\vspace{1mm}
\item	The detectron moves in space-time (as a function of $a$) slowly enough so that its field amplitude does not change significantly over a-intervals of length $M\delta$, where $M$ is an integer $>> 1$;
\vspace{1mm}
\item  The detectron moves in such a way that it uniformly samples the space-time extent of the detector. 
\vspace{1mm}
\item	The detectron becomes a detection when its oscillator's amplitude exceeds a fixed threshold $\Theta$.
\end{itemize}
We can apply this model to the two-slit setup shown in Figure 2.  The detectron moves within the space-time confines of the detection screen. The complex field amplitude $\psi(\mathbf{r},t)$ corresponds to the conventional Schr\"{o}dinger wavefunction at the screen, which in the ray approximation is given by:
\begin{equation}
\psi(0,L,z,t) \propto d_1^{-1} e^{i(k'd_1-\omega' t)}+ d_2^{-1} e^{i(k'd_2-\omega' t)}  \label{eq:1.11}
\end{equation}
where $k'$, $\omega'$ are the (observable) wave number and frequency, and $d_1 =(L_1^2+h^2)^{1/2} + (L_2^2+(z-h)^2)^{1/2}$, $d_2 =(L_1^2+h^2)^{1/2} + (L_2^2+(z+h)^2)^{1/2}$.
\begin{figure}
  \includegraphics[width=3in]{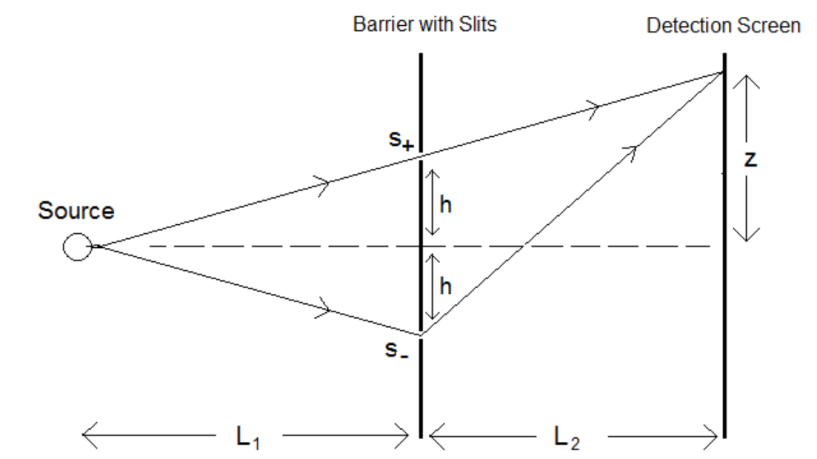}
\caption{Notation for quantum two-slit experiment}
\label{fig:2} 
\end{figure}
We simulated this system using MATLAB, with the parameters shown in Table 1 . We only considered a single time slice, and restricted to the $x=0$ portion of the screen.  The $z$ locations were discretized into 100 bins; the detectron jumped uniformly randomly from bin to bin every $M=400$ iteration steps.  At each iteration, the signal was incremented by $\nu_n \cdot \psi(0,L,z,0)$, where $\nu_n$ are i.i.d. complex random variables with standard normal real and imaginary parts. Each time the detection threshold $\Theta=500$ was reached, a detection was logged and the simulation was restarted. Altogether 100,000 detections were logged. Figure 3 shows the detection probability distribution obtained in the simulation. Agreement is very close with the theoretical result $|\psi(z)|^2$, with $\psi$ given by (\ref{eq:1.11}).  Note that in the simulation, distribution peaks are slightly lower than theoretical values. If quantum probabilities are indeed the result of such an accumulation process, it is possible that measured probability values may be lower than the conventional quantum prediction. Unfortunately, it is not possible to predict the extent of the lowering from our model, because it depends on details of the accumulation process that are not accessible to measurement. 
\begin{figure}[htb]
	   \includegraphics[width=4in]
	      {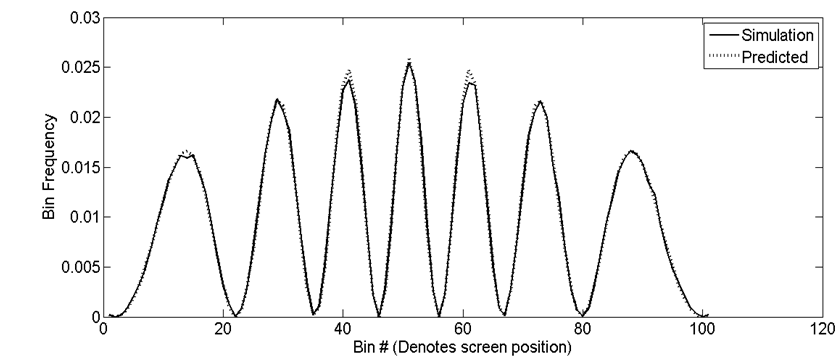}
	  \caption{Simulation and theory for double-slit experiment with the ray approximation}\label{fig:3}
\end{figure}
\begin{table}
\caption{Parameters for double-slit simulation with the ray approximation}
\label{tab:1}       
\begin{tabular}{lll}
\hline\noalign{\smallskip}
Parameter symbol& Parameter signifigance (distances in wavelengths)&Value \\
\noalign{\smallskip}\hline\noalign{\smallskip}
$h$ & 1/2 the distance between slits & 5\\ 
$L_1$ & Distance from source to slit screen & $1 \times 10^4$\\
$L_2$ & Distance from slit screen to detection screen & $1 \times 10^6$ \\ 
$M$ & Number from slit screen to detection screen & 400\\
$N_{\textrm{detect}}$ & Number of detections & $1 \times 10^5$\\
$Z$ & Screen half-width & $1 \times 10^6$\\
$z_n$ & Number of bins (discretization) & 100\\
$\Theta$ & Detection threshold & 500\\
\noalign{\smallskip}\hline
\end{tabular}
\end{table}
\section{Wavefunction formation via accumulation}
\label{sec:accumulation}
The preceding section describes how to obtain quantum-like detection probabilities given that a certain ``broadcast field" is present. However, it provides no mechanism for the creation of the broadcast field itself. In this section, we show how the broadcast field can be modeled as the result of a process of accumulation that parallels the signal accumulation described above.
The well-known path-integral expression for the propagator $K(\mathbf{r}_1,t_1;\mathbf{r}_2,t_2)$ is given by (\cite{Kleinert}):
\begin{equation}
K(\mathbf{r}_1,t_1;\mathbf{r}_2,t_2)=\int Dq(t) e^{iS[q(t)]/\hbar}   \label{eq:1.12}
\end{equation}
Here $S[q(t)]$ is the action, and the notation $\int Dq(t)$ denotes an equally-weighted summation over all possible paths from $(\mathbf{r}_1,t_1)$ to $(\mathbf{r}_2,t_2)$.
This integral may be seen as the outcome of an accumulation process.  We may envision a succession of carrier-wave \emph{blips}, where each blip corresponds to a single path $q(t)$ and makes a differential contribution to the field $\psi(\mathbf{r}_2,t_2)$ which is proportional to $e^{iS[q(t)]/\hbar}\psi(\mathbf{r}_1,t_1)$ (as shown in Figure 4).  Recall that $\psi(\mathbf{r},t)$ corresponds to the amplitude and phase of a modulated carrier wave; thus $e^{iS[q(t)]/\hbar}$ expresses the influence of a source at $(\mathbf{r}_1,t_1)$ on the amplitude and phase of the wave at $(\mathbf{r}_2,t_2)$ when a blip passes between them. 
\begin{figure}[htb]
	   \includegraphics[width=4in]
	      {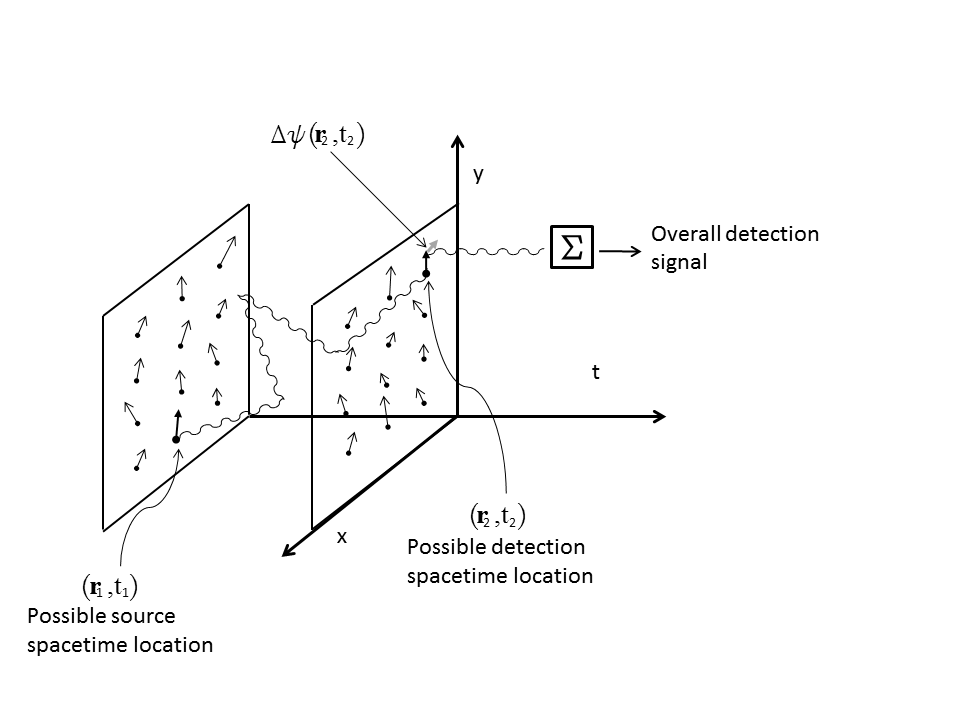}
	  \caption{Single ``blip" with field perturbation at $(\mathbf{r}_2,t_2)$}\label{fig:4}
\end{figure}

\begin{figure}[htb]
	\includegraphics[width=4in]
	      {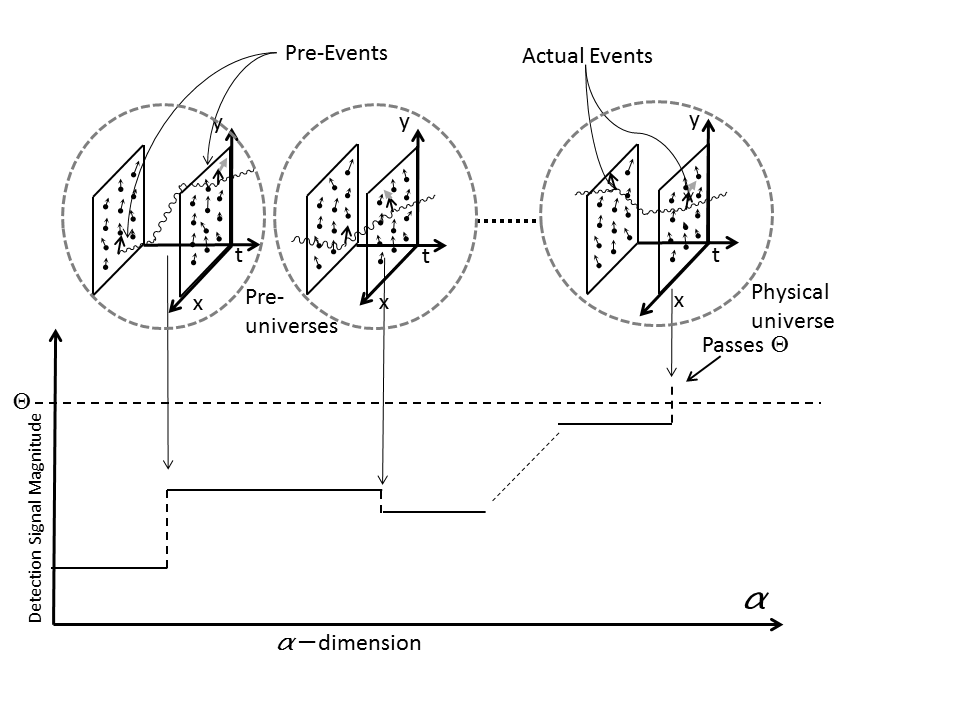}
	  \caption{Detection signal buildup and detection}\label{fig:5}
\end{figure}

These blips may be associated with detectrons as follows.  If $(\mathbf{r}_2,t_2)$  is a possible event detection location, then a path $q(t)$ that passes through $(\mathbf{r}_2,t_2)$  can be identified with a detectron location of $(\mathbf{r}_2,t_2)$. We postulate that simultaneously with causing an incremental change in the field $\psi(\mathbf{r}_2,t_2)$, the blip also increments the overall complex detection signal amplitude by $\psi(\mathbf{r}_2,t_2) \cdot \sum_n \nu_n$, as described in the communication model in the previous section.  In this way, the blips perform a dual mathematical function: they both build up the field, and furnish the uniform random sampling of possible detection sites that is required to obtain Born-rule probabilities (as was shown in the previous section). The process of detection signal buildup and detection is shown in Figure~\ref{fig:5}.

To see how this works in practice, we focus specifically on the two-slit experiment shown in Figure 2 in two space dimensions, using non-relativistic electrons of fixed energy $\hbar \omega'$ as particles.  We consider the spatial distribution of detections at time $t_2=0$; due to invariance, this distribution will be independent of $t_2$. The detection screen corresponds to spatial locations $(L_2,z)$. We assume the source is configured so that $\psi(s_{+},t)=\psi(s_{-},t)= Ae^{-i\omega' t}$,  where $s_{+}$ and $s_{-}$ denote the spatial locations of the two slits ( $(0,h)$ and $(0,-h)$, respectively). We also assume that $A>0$ is large enough so that the signal accumulation process has negligible effect on the size of $A$. We replace $t$ with negative $t$ (since only negative times contribute to detection at $t_2=0$) and obtain an expression the Schr\"{o}dinger kernel for paths that exit through the upper slit $s_+$:
\begin{equation}
K(s_+,-t;\mathbf{r}_2,0) \equiv \kappa(z-h,t) \equiv (B/t) \cdot \exp\left(\frac{i(L_2^2+(z-h)^2))}{2 \hbar t }\right)  \label{eq:1.13}
\end{equation}
where $B$ is a constant of proportionality. Similarly, the kernel for paths that exit through the lower slit is $K(s_-,-t;\mathbf{r}_2,0)=\kappa(z+h,t)$.
The theoretical expression for the wavefunction $\psi(z,0)$ is
\begin{equation}
\psi(z,0) \propto \int_0^{\infty} \kappa(z-h,t) e^{i\omega't} dt+\int_0^{\infty} \kappa(z+h,t) e^{i \omega't} dt   \label{eq:1.14}
\end{equation}
which evaluates to
\begin{equation}
\psi(z,0) \propto K_0 \left(-2i \sqrt{\beta_{-} \omega'}\right) + K_0 \left(-2i \sqrt{\beta_{+} \omega'}\right)  \label{eq:1.15}
\end{equation}
where
\begin{equation}
\beta_{\pm}=\frac{L_2^2+(z \pm h)^2)}{2\hbar/m}.   \label{eq:1.16}
\end{equation}
It was computationally intractable for us to simulate the path integrals that give rise to the Schr\"{o}edinger kernels themselves. Instead, we assumed the kernels and simulated the cumulative effect of different source points. We use the following algorithm to accumulate both the fields at different screen locations and the overall detection signal:
\vspace{1mm}

\begin{samepage}
\noindent\texttt{Initialize:} $\psi(z)=0$ for all detection locations $z; n=0$\\
\texttt{While} $|\textrm{Signal}|<\Theta$\\ 
\indent $n = n+1$\\
\indent\texttt{If } $n$ divides $m$\\
\indent \indent Change current detectron location $z_n$;\\
\indent\indent Choose random time $t$ (uniformly distributed);\\
\indent \texttt{End If}\\
\indent $\psi(z_n)=\psi(z_n)+ \kappa(z_n \pm h,t)\cdot e^{i\omega't}$;\\		
\indent $\nu= \textrm{Normal}(0,1)+ i\cdot \textrm{Normal}(0,1)$;\\
\indent\textrm{Signal=Signal + $\nu \cdot \psi(z_n)$;}\\
\texttt{End While}\\
Record detection at location $z_n$\\
\end{samepage}

\begin{figure}[htb]
	\includegraphics[width=5in]
	      {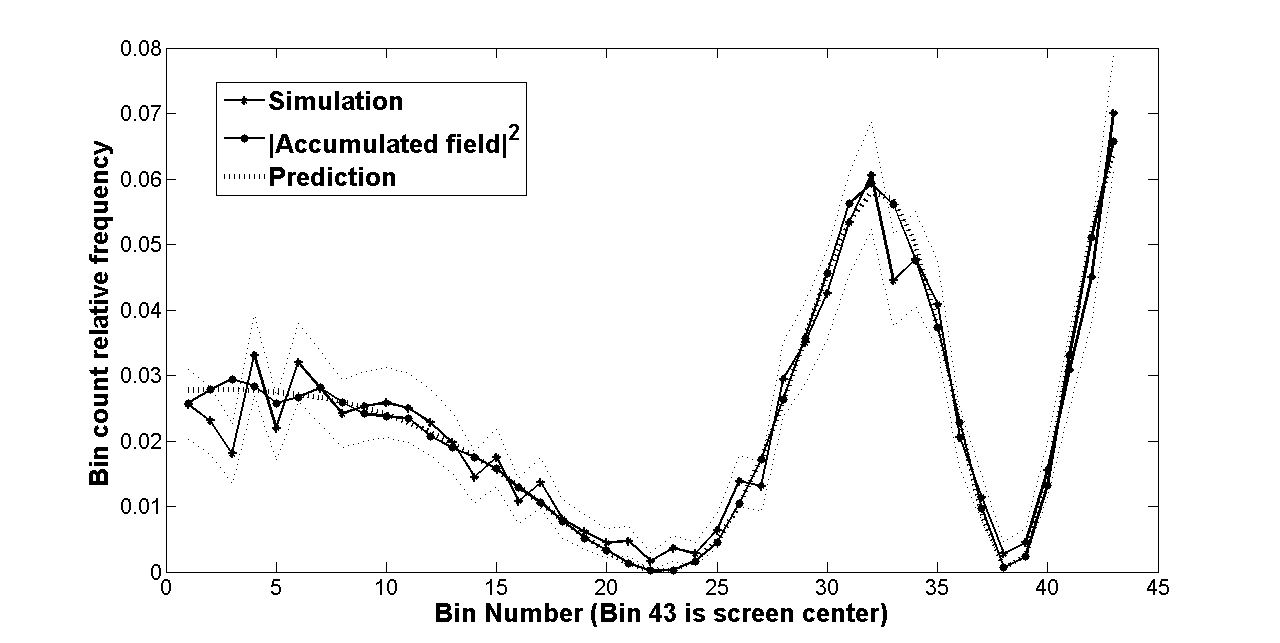}
	  \caption{Accumulated field and detection frequencies for 2-slit experiment}\label{fig:6}
\end{figure}

\begin{table}
\caption{Parameters for 2-dimensional double slit with non-relativistic electrons}
\label{tab:2}       
\begin{tabular}{lll}
\hline\noalign{\smallskip}
Parameter symbol& Description &Value (in mks units)\\
\noalign{\smallskip}\hline\noalign{\smallskip}
$h$ & 1/2 the distance between slits & $7.26 \times 10^{-10}$ m\\ 
$L_2$ & Distance from slits to detection screen & $7.26 \times 10^{-9}$ m \\ 
$M$ & Number of iterations between detectron jumps & 25\\
$N_{\textrm{detect}}$ & Number of detections & 3600\\
$T_{\textrm{min}}$ & Minimum value for random time $t$ & $3.37 \times 10^{-15}$ m sec\\
$T_{\textrm{max}}$ & Maximum value for random time $t$ & $3.02 \times 10^{-14}$ sec\\
$\omega'$ & Electron frequency & $5 \times 10^{15}$ Hz\\
$Z$ & Screen half-width & $1.45 \times 10^{-8}$ m \\
$z_n$ & Number of bins (discretization) & 43\\
$\Theta$ & Detection threshold & $1\times 10^5$\\
\noalign{\smallskip}\hline
\end{tabular}
\end{table}
Table 2 shows the parameters of the simulation.  Figure~\ref{fig:6} shows the results for 3600 detections. The figure shows counts from 1/2 of the detection screen ($z<0$), which was discretized into 43 bins (with the edge of bin 43 at the center of the screen). The simulation relative counts per bin and the accumulated field $|\psi(z)|^2$ are plotted, as well as theoretical detection probabilities from (\ref{eq:1.15}). In general, the theoretical curve lies within error bars of the simulation counts; the simulation counts are consistently slightly higher than theoretical probabilities near the interference pattern nulls. 
\section{Multiple quantum events}\label{sec:multiple}
In the preceding sections we have described how accumulation of the signal associated with the moving detectron eventually leads to detection when the accumulated signal passes a threshold. This description treats a single quantum event in isolation without considering interactions between quantum events. Accordingly, we now postulate that every space-time event corresponds to a detectron, and all detectron signals at a given $a$-instant are \emph{multiplied} before accumulation.  The situation thus remains as shown in Figure~\ref{fig:5}, except that the jumps in detection signal magnitude are caused by products of detectron signals. In fact, the effects of this multiplication are already included in the model in Section~\ref{sec:single} via the  random Gaussian factors ${\nu_n}$  which multiply the field $\psi(\mathbf{r},t)$ at the detectron: these fluctuations in the signal amplitude correspond to the variation of other detectron signals.\footnote{The actual distribution arising from a product of random signals will be lognormal rather than normal: we used normal random variables for computational simplicity. The results are not sensitive to the particular shape of the distribution.}

Along with the change in detection signal,  at each $a$-instant the complex amplitude of the field at each detectron location is incremented as follows. Let $(\mathbf{r}_1,t_1), \ldots (\mathbf{r}_M,t_M)$ be the (time-ordered) sequence detectron space-time locations for the pre-universe at $a$. Then the complex wavefunction amplitude at $(\mathbf{r}_m,t_m)$ is incremented by $\prod_{j<m} e^{i S[q_{j,j+1}(t)]/\hbar}$, where $q_{j,j+1}(t)$ is a path from $(\mathbf{r}_j,t_j)$ to $(\mathbf{r}_{j+1},t_{j+1})$. Given that the paths $q_{j_1,j_1+1}(t)$ and $q_{j_2,j_2+1}(t)$ are statistically independent for $j_1 \neq j_2$, this gives rise to the expression (\ref{eq:1.12}) for the propagator.

In summary, each pre-universe shown in Figure~\ref{fig:5} gives a single contribution to the overall detection signal, that accumulates as the pre-universes unfold in the $a$-dimension. Once the detection signal attains a threshold for a particular value of $a$, the space-time universe that we experience is actualized as a ``snapshot'' at that $a$-instant.

\section{Extension to a field theory}\label{sec:extension}
In the previous sections, we have presented detectrons as moving placeholders for possible detection sites. This picture presumes a division of the universe between fields and detectors.  This division of course is not realistic, for the detectors themselves are represented by fields in their own right. We may eliminate this dichotomy by identifying detectrons as field configurations (including the transmitted particle field and the field representing the atoms of the detection screen) that produce contributions to the detection signal.
The transition to a field theory may then be accomplished by replacing paths with field configurations. In other words, the pre-universe at each $a$-instant has a particular field configuration that produces an overall contribution to the detection signal. As $a$ increases, the field configurations vary and their contributions to the detection signal are accumulated. At the $a$-instant where the threshold is attained, we arrive at the field configuration corresponding to the observable universe. 

\section{Explanations of quantum paradoxes}
\emph{Collapse of the wave packet}:
In our model, the wave function is seen as an ``actual" field that develops via a process of accumulation. The field is not merely a representation of the observer's partial knowledge; but on the other hand, the field is not directly observable via physical events in space-time. The apparent ``collapse'' of the wave packet is due to the fact that the physical universe is only a single ``$a$-slice'' of the entire process. 
\vspace{1mm}

\noindent\emph{EPR and Bell's inequality}:
Non-local effects pose no problem for this model, for the model itself is inherently non-local. An EPR experiment where the two spin detectors are aligned parallel will always detect anti-aligned particles because both detections correspond to the same blip and are thus perfectly correlated.  If the detectors are not parallel, the detection probabilities are still determined according to the quantum expression appropriate for that configuration.
\vspace{1mm}

\noindent
\emph{Aharanov-Bohm Effect}:
The Aharanov-Bohm effect shows that fields that are localized in a region where a particle can never be detected can still have an effect on the motion of the particle. This poses no difficulty to our model, because in our model the so-called particle is not an object that travels through space-time but  rather a correlated series of detection events. 
\vspace{1mm}

\noindent
\emph{Identical particle statistics}: 
More work is needed to introduce spin into the model. However, in light of our framework it is not surprising that ``particles" obey special statistics, because ``particles" are not separate objects at all. What we call a ``particle" is simply a series of correlated events. 
\section{Comparison With Other Interpretations of Quantum Mechanics}
We briefly compare our interpretation with other alternative interpretations of quantum mechanics.
\vspace{1mm}

\noindent
\emph{Everett's ``Many-worlds" interpretation}\cite{DeWitt} requires exponential plethorization of space-times. Our model, which embeds space-time within one additional dimension, possesses a much simpler state space.
\vspace{1mm}

\noindent
\emph{Bohm's quantum mechanics}\cite{Bohm} posits that particles such as electrons are able to track along with pilot waves. This appears to imply that these particles have some sort of inner structure. In our model particles are not ``objects" at all, so no such complications appear. 
\vspace{1mm}

\noindent
\emph{Cramer's transactional quantum mechanics}\cite{Cramer} interprets $\psi^*$ as a wave traveling backwards in time,  but gives no explanation why $\psi\psi^*$ should be interpreted as a probability. Furthermore, transactional quantum mechanics is not very clear about the order in which ``transactions" are determined. In our model, all transactions are determined ``simultaneously" (at $a=a_\Theta)$, and a single accumulation process is used to determine all interaction events.
\vspace{1mm}

\noindent
We also remark that none of these alternative models explains why the wavefunction is complex, nor why the squared amplitude is interpreted as a probability.
\section{Possible experimental verification}
If true, then our model indicates that the usual formula for a quantum wavefunction is a statistical approximation, and small deviations from the probabilities predicted by the wave quation should be expected. In particular, in our simulations we consistently found that detection rates near theoretical wavefunction nulls were higher that the conventional quantum predictions.  Unfortunately, the size of these effects would depend on aspects of the process that cannot be directly measured.
\section{Conclusions}
Our model presents a radically different picture of reality. Traveling ``particles" are replaced with series of detection events; as a visual analogy, imagine a series of fireflies in a line that flash successively, giving the impression that a single firefly is moving along the line.
Our model replaces temporal causality with atemporal causality; past, present, future are actualized together as the result of a process that occurs in a different dimension. Apparent temporal causality is due to correlation and not causation.
The wavefunction is given a physical interpretation as a dynamical field; and the Born rule based on the wavefunction is derived as the natural result of a thresholding process involving this field.
The model includes possible differences from conventional quantum mechanics. A lowering of peak probabilities in quantum interference patterns compared to the conventional quantum-mechanical prediction is a possible result of the model. 

\section{Appendix: Mathematical derivation of the Born rule}
In this section, we prove the Born probability rule,
\begin{equation}
\Pr\left[\kappa \left( \left \lceil N_{\Theta} / M \right \rceil \right)=k\right] \propto |\psi_k |^2,	\qquad k =1,\ldots ,K,   \label {eq:1.17}
\end{equation}
 for the  wireless communication scenario described in Section \ref{sec:wireless}. We will use the notation and definitions of that section. 
 
In order to investigate the dependence of $Pr\left[\kappa \left( \left \lceil N_{\Theta} / M \right \rceil \right)=k\right]$ on $\psi_k$, for each fixed $m'>0$ we will investigate the event 
\begin{equation}
E_{m',k} \equiv\left[ m'=\left \lceil  N_{\Theta} / M \right \rceil \text{~and~} \kappa(m' )=k\right]
\end{equation}
conditioned on fixed sequences  of $(m'-1)$ initial $\psi'$s, corresponding to the $K^{m'-1}$ events
\begin{align}
&F_{m'}(\lbrace k'_{1},\ldots k'_{m'-1}\rbrace) \equiv \lbrace \kappa(m) = k'_m , 1\le m<m',\notag\\
&\qquad{} \textrm{where} ~ 1\le k'_m \le K ~ \textrm{are~ fixed}\rbrace  \label{eq:1.18}
\end{align}
 We shall show that 
\begin{equation}
\Pr[E_{m',k} | F_m'(\lbrace k'_{1},\ldots k'_{m'-1}\rbrace)]= C(m',\lbrace k'_{1'},\ldots k'_{m'-1}\rbrace) \cdot|\psi_k |^2,   \label{eq:1.19}  
\end{equation}
where $C(\ldots )$ is independent of $k$. The events $ \lbrace F_m'(\lbrace k'_{1},\ldots k'_{m'-1}\rbrace)\rbrace $ for fixed $m'$ partition the sample space and $\Pr[F_{m'}(\lbrace k'_{1},\ldots k'_{m'-1}\rbrace)] = K^{1-m'}$. Furthermore, the events $\left\{E_{m',k} \right\}_{m'=1,2,\ldots }$ partition the event $\left\{\left[\kappa \left( \left \lceil N_{\Theta} / M \right \rceil \right)=k\right]\right\}$, so we obtain 
\begin{align}
&\Pr\left[\kappa \left( \left \lceil N_{\Theta} / M \right \rceil \right)=k\right]     \notag\\
&= \sum_{m'}\sum_{k'_{1},\ldots ,k'_{m'}}\Pr[E_{m',k} | F_{m'} (\lbrace k'_{1},\ldots k'_{m'-1}\rbrace)]\Pr[F_{m'} (\lbrace k'_{1},\ldots k'_{m'-1}\rbrace)]   \label{eq:1.20}\\
&= |\psi_k |^2 \sum_{m'}\sum_{k'_{1},\ldots ,k'_{m'}}C(m',k'_1,\ldots k'_{m'-1}) \cdot K^{1-m'}   \notag\\
&\propto |\psi_k |^2  \notag
\end{align}
which is the desired result.

We prove (\ref{eq:1.19}) as follows. Conditioned on event  $F_m'(\lbrace k'_{1},\ldots k'_{m'-1}\rbrace)$ , we have 
\begin{equation}
S(N)= \sum_{n=1}^N(\psi_{k'_{\lceil n/M \rceil}} \cdot \nu_n),\qquad (N \le m'M)   \label{eq:1.22}
\end{equation}
were the $\lbrace \nu_n \rbrace$ have i.i.d. standard normal real and imaginary parts (we write this as: $\nu_n \tilde N(0,1) + iN(0,1)$).  It follows that $S(N)$ is a random walk in the complex plane with independent (but not identically distributed) steps.  We also have
\begin{equation}
E[|S(N)|^2 ]=2 \sum_{n=1}^N \left|\psi_{k'_{\lceil n/M \rceil }} \right|^2  \label{eq:1.23}
\end{equation}
We shall assume that $\Theta >> \max_k⁡|\psi_k |$, so $\Theta$ is much greater than any individual term in $S(N)$. It follows that the distribution of $\lbrace \Theta^{-1}S(N)\rbrace$ for all sample paths $S$ can be approximated as a standard Brownian motion $B(\tau)$, where the time variable $\tau$ is given by  
\begin{equation}
\tau(N) \equiv E \left[|\Theta^{-1} S(N)|^2 \right] \approx 2\Theta^{-2} \sum_{n=1}^N \left|\psi_{k'_{\lceil n/M \rceil}} \right|^2 .  \label{eq:1.24}
\end{equation}
The sample paths comprised in the event $ \left \lceil N_{\Theta} / M \right \rceil \ge m' $ correspond in the Brownian motion picture to sample paths for which $|B(t)| <1$ for all $t \le \tau((m'-1)M)$.  For these sample paths, the distribution of $B(\tau((m'-1)M))$ corresponds to the position probability density for a standard Brownian motion with absorbing barrier at $|z|=1$.   

Now there is a close connection between Brownian motion and the heat equation as follows. Let $\beta(z,T)$  be the probability density at time $T$  of a Brownian motion with absorbing barrier at $|z|=1$. Then $\beta(z,T)$  can be found by solving the heat equation with corresponding boundary and initial conditions, which in this case are:
\begin{description}
\item[] Boundary conditions:  $\beta(z,T) = 0$ for $|z|=1$;
\item[] Initial conditions: $\beta(z,0)=\delta(z)$,
\end{description}
where $\delta$(\ldots ) is the Dirac delta function. We do not need the complete solution for $\beta(z,T)$ (which can be expressed in terms of the Bessel functions $\lbrace J_0(\alpha_n r/\Theta)\rbrace n=1,2,\ldots )$, but we will make use of the following properties:
\begin{description}
\item[a)]	$\beta$ is radial, so we may write $\beta(z,T)$ as $\beta(r,T)$
\vspace{1mm}
\item[b)]	$\beta(r,T)$ is $C^{\infty}$ for $r \le \Theta$ and  $T > 0$;
\vspace{1mm}
\item[c)]       $\beta_r(1,T) < 0$ for all $T > 0$;
\end{description}
These properties can be mathematically proven, but are also intuitive consequences of the physical interpretation of $\beta(r,T)$ as an evolving temperature distribution within a disk where the boundary is held at zero temperature.
In light of property c), at  time $T \beta(r,T)$ can be approximated near the boundary $|z| = 1$ as
\begin{equation} 
\beta(r,T) = (1-r)|\beta_r(1,T)| + O[(1-r)^2].   \label{eq:1.25}
\end{equation} 
It follows from our identification of $\lbrace S(N)/\Theta \rbrace$ with $B(\tau(N))$ that
\begin{equation}
dP \left[|S((m'-1) \cdot M)| = r\Theta ~ \textrm{and} ~ \left \lceil N_{\Theta} / M \right \rceil \ge m' \right]= A(1-r) +  O(1-r)^2.    \label{eq:1.26}
\end{equation}
Since $\Psi_m=\psi_k$,it follows that the terms $\Psi_{\lceil n/M \rceil}  \cdot \nu_n \sim |\psi_k | \cdot [N(0,1)+i \cdot N(0,1)]$ for $n=(m'-1)M+1\ldots m'M$. By rotating in the complex plane we have that 
\begin{align}
&\Pr \left[|S(n)|<\Theta,n=(m'-1)M+1 \ldots m'M \right. \notag \\
&\qquad \qquad \left| \Psi_{m'}=\psi_k ~\textrm{and}~ |S((m'-1)M)|=r\Theta \right]     \label{eq:1.27}\\
&= \Pr \left[ \left|r+|\psi_k/\Theta| \sum_{j=1 \ldots J} \nu'_j   )\right|<1,J=1 \ldots M \right],  \notag
\end{align}
where  $\nu'_j \sim N(0,1)+i \cdot N(0,1)$. 

In the case where $\Theta>> \max_k |\psi_k|$ and $r\approx 1$, we have
\begin{equation}
	\left| r+|\psi_k/\Theta| \sum_{j=1 \ldots J} \nu'_j \right| = \Re\left[r+|\psi_k/\Theta| \sum_{j=1 \ldots J} \nu'_j\right] + O([|\max_k |\psi_k| / \Theta]^2).  \label{eq:1.28}
\end{equation}
Thus the condition $\left| r+|\psi_k/\Theta| \sum_{j=1 \ldots J} \nu'_j \right|<1$ can be replaced to a very close approximation by the condition 
$\Re \left[ r+|\psi_k/\Theta| \sum_{j=1 \ldots J} \nu'_j \right]< 1$ (see also Figure 7) and 
\begin{align}
&\Pr[|S(n)|<\Theta,n=(m'-1)M+1 \ldots m'M  \notag\\
&\qquad{} | \Psi_{m'}=\psi_k \textrm{and}~ |S((m'-1)M)|=r\Theta ]     \label{eq:1.29}\\
&\approx \Pr \left[\Re\left[r+|\psi_k/\Theta| \sum_{j=1 \ldots J} \nu'_j \right]<1,J=1 \ldots M\right]   \notag\\
&=  \Pr\left[\Re\left[\sum_{j=1 \ldots J} \nu'_j\right]<\Theta(1-r)/|\psi_k |,J=1\ldots M\right]    \notag\\
&\equiv \phi \left(\frac{\Theta(1-r)}{|\psi_k |} \right).   \notag
\end{align}
\begin{figure}[htb]
\includegraphics[width=2.5in]
{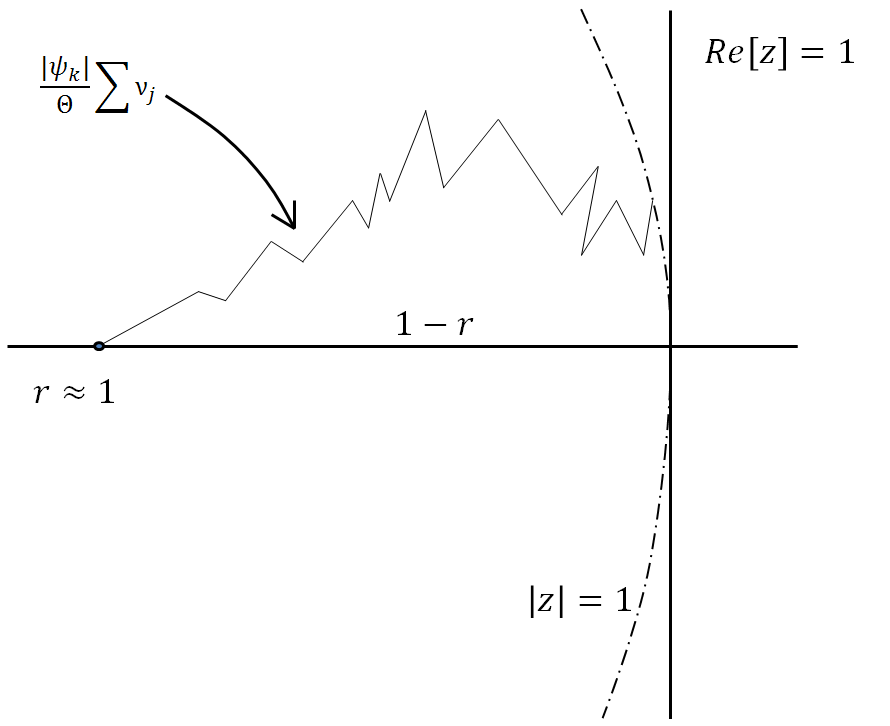}
\caption{Depiction of random sum}\label{fig:7}
\end{figure}
Note that $\phi(x)=1$ for $x\ge J$ since $\Re[|\nu'_j|] \le 1$.
Now for sample paths with $\left \lceil N_{\Theta}/M \right \rceil \ge m',$ the event $\left \lceil N_{\Theta}/M \right \rceil = m',$ is the complement of the event $\lbrace |S(n)|< \Theta,n=(m'-1)M+1 \ldots m'M \rbrace$. It follows in summary that
\begin{align}
&\Pr[E_{m',k} | Fm'(\lbrace k'_1,\ldots k'_{m'-1}\rbrace)]   \notag\\
&=\Pr\left[ \lceil N_{\Theta} / M \rceil = m' ~\text{and}~  \Psi_m=\psi_k \right. \left| F_{m'} (\lbrace k'_{1},\ldots k'_{m'-1}\rbrace)\right]   \notag\\
&= \Pr\left[ \left( \lceil N_{\Theta} / M \rceil  = m' |  \Psi_m=\psi_k \right) \right. \left| F_{m'} (\lbrace k'_{1},\ldots k'_{m'-1}\rbrace)\right]\cdot \Pr[\Psi_m=\psi_k]     \notag\\
&\equiv \int_0^1 \left(1-\phi \left(\frac{\Theta(1-r)}{|\psi_k|}\right)\right) \notag\\
&\qquad \qquad \cdot dP \left[|S((m'-1)\cdot M)|= r\Theta \textrm{~and}~  \lceil N_{\Theta}/M \rceil \ge m' \right] \cdot K^{-1}.   \label{eq:1.30}
\end{align}

Note that $(1-\phi(\Theta(1-r)/|\psi_k |))=0$ unless $0< (1-r) < J \cdot |\psi_k |/\Theta$; and since $\Theta >> J \cdot |\psi_k|$, the approximation (\ref{eq:1.26}) holds on this range. Our integral becomes 
\begin{equation}
\approx \int_{1-J|\psi_k |/\Theta}^1 \left(1-\phi\left(\frac{\Theta(1-r)}{|\psi_k |}\right)\right) \cdot K^{-1} A(1-r)\cdot dr.  \label{eq:1.31}
\end{equation}
Changing variable to $x \equiv(1-r) / |\psi_k |$, we have 

\begin{align}
&\approx \int_0^{J/\Theta} (1-\phi(x \Theta)) \cdot |\psi_k |^2 \cdot K^{-1} Ax \cdot dx \label{eq:1.32}\\
&\propto |\psi_k |^2.    \notag
\end{align}


\begin{acknowledgements}
Thanks to Walter Wilcox for many helpful suggestions.
\end{acknowledgements}



\end{document}